\documentstyle[psfig]{europhys}

\def\etal{{\hbox{{\tenit\ et al.\/}\tenrm :\ }}}

\def\And{{\rm and\ }}

\def\stars{\bigskip\centerline{***}\medskip}

\newif\ifboo \boofalse

\def\Review#1{\boofalse{\it #1},}
\def\Name#1{{\sc #1},}
\def\Vol#1{\ifboo Vol. {\bf #1}\else{\bf #1}\fi}
\def\Year#1{\ifboo #1\else(#1)\fi}

\def\Page#1{\ifboo {\rm p. #1}\else{\rm #1}\fi}

\begin{document}
\euro{xx}{x}{xx-xx}{xxxx}
\Date{}
\shorttitle{C. GRIMALDI: SPIN-ORBIT SCATTERING IN
ETC.}
\title{{\Large Spin-orbit scattering in $d$-wave superconductors}} 
\author{C. Grimaldi} 
\institute{ \'Ecole 
Polytechnique F\'ed\'erale de Lausanne,
DMT-IPM, CH-1015 Lausanne, Switzerland}
\rec{}{}
\pacs{
\Pacs{74}{20$Fg$}{BCS theory and its development}
\Pacs{71}{70$Ej$}{Spin-orbit coupling, Zeeman and Stark splitting}
\Pacs{74}{62$Dh$}{Effects of crystal defects, doping and substitution}
}

\maketitle 
\begin{abstract}
When non-magnetic impurities are introduced in a $d$-wave
superconductor, both thermodynamic and spectral properties
are strongly affected if the impurity potential is close
to the strong resonance limit. In addition to the scalar impurity
potential, the charge carriers are also spin-orbit coupled to the impurities.
Here it is shown that 
(i) close to the unitarity limit for the impurity scattering, the
spin-orbit contribution is of the same order of magnitude than the
scalar scattering and cannot be neglected, 
(ii) the spin-orbit scattering is pair-breaking 
and (iii) induces a small $id_{xy}$ component to the off-diagonal 
part of the self-energy.   
\end{abstract}

In high-$T_c$ superconductors, disorder has 
important effects on both thermodynamic and spectral properties.
The critical temperature $T_c$ and the superfluid density $\rho_s$
are lowered by non-magnetic impurity substitution 
\cite{hardy,ulm,bernhard,nachumi} and disorder
induced by irradiation \cite{tolpygo,vobornik}. Recent ARPES data
show clearly how disorder leads to a redistribution of spectral
intensity by adding new states at the Fermi level \cite{vobornik}.
The basic elements of the current theory have been inspired by
previous studies on heavy fermion superconductors and are given
by the anisotropy of the order parameter and the strong resonance
limit for the impurity potential \cite{pines,hirsch1}. 
These elements, adjusted to describe condensates with a $d$-wave
symmetry of the order parameter, are able to account for
most of the features observed by experiments on
high-$T_c$ $d$-wave superconductors \cite{hirsch2,kim,maki1}.
However, discrepancies still exist, like the overestimation
of the $T_c$-suppression \cite{franz}. In order to correct this
situation, and to provide a more realistic picture,
several improvements of the theory have been proposed 
\cite{radtke,carbotte,fehren}, and, recently, the effect of spatial variation
of the order parameter has been taken into account \cite{franz,zhito,hirsch3}.

In addition to the scalar impurity potential, the charge carriers
are also spin-orbit coupled to the impurities. So far, this additional
scattering channel has not been considered because the spin-orbit
interaction is believed to provide, if any, only negligible effects
(at least in the absence of a Zeeman magnetic field). This argument is based
on the observation that the spin-orbit potential is of order
$v_{so}\sim \Delta g\, v$ where $v$ is the impurity
potential and $\Delta g$ is the shift of the $g$-factor \cite{elliott}. 
The value of $\Delta g$ depends on the specific impurity, however
its order of magnitude is roughly $\Delta g\simeq 0.1$. 
From this estimate, it is expected therefore that the spin-orbit 
scattering rate $1/\tau_{so}\simeq 
N_0v_{so}^2$, where $N_0$ is the charge carriers density of states,
should be at most of order $1/\tau_{so}\sim 10^{-2}/\tau_{imp}$, 
therefore  negligible with respect to the scalar impurity scattering rate 
$1/\tau_{imp}$ \cite{yang}.
Although such an estimate is correct in the
Born approximation (weak scattering) nevertheless it underestimates 
the effect by orders of magnitude in the strong resonance limit, believed
to be valid for high-$T_c$ superconductors.
To illustrate such a substantial discrepancy between the Born and
the unitarity limit, let us first consider the self-consistent
t-matrix solution of the impurity problem for $\Delta g=0$. In
this case the electron (hole)
propagator is $G^{-1}({\bf k},i\omega_n)
=i\tilde{\omega}_n-\rho_3\epsilon({\bf k})-
\rho_1\Delta({\bf k})$, where $\Delta({\bf k})=\Delta\cos(2\phi)$ 
is the $d$-wave order parameter and $\phi$ is the polar angle,
$\epsilon({\bf k})$ is the electron dispersion for an half-filled band
and $\rho_1$, $\rho_3$ are Pauli matrices.
The renormalized Matsubara frequency $i\tilde{\omega}_n$ satisfies
the following equation \cite{hirsch1,maki1}:
\begin{equation}
\label{renorm}
i\tilde{\omega}_n=i\omega_n+\Gamma\frac{g_0(i\omega_n)}
{c^2-g_0(i\omega_n)^2},
\end{equation}
where $g_0(i\omega_n)=
\langle i\tilde{\omega}_n/[\tilde{\omega}_n^2+\Delta({\bf k})^2]^{1/2}\rangle$
and $\langle\cdots\rangle$ is the average over the polar angle $\phi$.
In equation (\ref{renorm}), $\Gamma=n_i/(\pi N_0)$
and $c=1/(\pi N_0 v)$ where $n_i$ is the impurity
concentration.
For $c\gg 1$, eq.(\ref{renorm}) reduces to the
Born limit while for $c\ll 1$ it leads to the unitarity, or 
strong resonant, limit.
Now, let us suppose that the spin-orbit impurity scattering leads
to a renormalization contribution of the same form of eq.(\ref{renorm}).
Since $v_{so}/v\sim \Delta g$, the renormalization induced by
both the impurity and the spin-orbit scatterings can be estimated by:
\begin{equation}
\label{renorm2}
i\tilde{\omega}_n=i\omega_n+\Gamma\frac{g_0(i\omega_n)}
{c^2-g_0(i\omega_n)^2}+\Gamma\frac{g_0(i\omega_n)}
{(c/\Delta g)^2-g_0(i\omega_n)^2}.
\end{equation}
For $c\gg 1$, $i\tilde{\omega}_m\simeq i\omega_n+
\Gamma g_0(i\omega_n)c^{-2}(1+\Delta g^2)$ and, as expected in the Born
limit, the contribution of spin-orbit
scattering is $\Delta g^2$ times smaller than that of scalar impurity 
scattering. 
On the other hand, for strong scattering, $c/\Delta g$ can be
very small and eventually it vanishes in the unitarity limit $c\rightarrow 0$. 
As a result, in this limit the spin-orbit scattering leads to
a renormalization of the same order of the impurity scattering, namely of
order $\Gamma/\Delta$.  

To provide more solid grounds to the above simple picture, 
it is necessary to treat the impurity and spin-orbit interactions 
on the same level by generalizing the usual
t-matrix approach for the impurity scattering also to the spin-orbit
contribution. To this end,
let us start by considering the spin-orbit impurity potential.
Two-dimensionality is often assumed in describing the main electronic
excitations at least for some of the high-$T_c$ superconductors. In the
present context, the reduced dimensionality has the following implication.
If the charge carriers are confined to move in the $x$-$y$ plane and
the impurity potential is  $V({\bf r})=v\sum_{i,{\bf k}}\exp[i{\bf k}\cdot
({\bf r}-{\bf R}_i)]$, 
where ${\bf R}_i$ denotes the impurity positions,
the spin-orbit interaction assumes the following form:
\begin{equation}
\label{spinpot}
V_{so}({\bf r})=i\eta_{so} v\sum_{i,{\bf k}}e^{i{\bf k}\cdot
({\bf r}-{\bf R}_i)}[{\bf k}\times{\bf p}]\cdot\mbox{\boldmath$\sigma$}=
i\Delta g\,v\sum_{i,{\bf k}}e^{i{\bf k}\cdot
({\bf r}-{\bf R}_i)}\frac{[{\bf k}\times{\bf p}]_z}{k_F^2}\sigma_z ,
\end{equation}
where ${\bf p}=-i\mbox{\boldmath$\nabla$}_{{\bf r}}$ is the momentum operator
and the spin-orbit coupling has been parametrized by $\eta_{so} v= 
\Delta g\,v/k_F^2$ where $k_F$ is the Fermi momentum. The effect of
two-dimensionality is therefore to couple the electron spin only along 
the $z$ direction. Hence, by choosing the $z$ axis as the direction of spin
quantization, the spin-orbit interaction (\ref{spinpot}) does not
mix the spin components. The generalized Green's
function in the particle-hole spin space is:
\begin{equation}
\label{green1}
G({\bf k},i\omega_n)^{-1}=G_0({\bf k},i\omega_n)^{-1}-
\Sigma({\bf k},i\omega_n),
\end{equation}
where $G_0({\bf k},i\omega_n)^{-1}=i\omega_n-\rho_3\epsilon({\bf k})
-\rho_2\tau_2\Delta({\bf k})$ and the Pauli matrices $\rho_i$ and $\tau_j$
act on the particle-hole and spin subspaces, respectively \cite{fulde}. In the 
self-consistent t-matrix approximation the self-energy is
$\Sigma({\bf k},i\omega_n)=n_iT_{tot}({\bf k},{\bf k},i\omega_n)$ where the 
t-matrix satisfies the following equation:
\begin{equation}
\label{tmatrix1}
T_{tot}({\bf k},{\bf k}',i\omega_n)=u({\bf k},{\bf k}')+\sum_{{\bf k}''}
u({\bf k},{\bf k}'')G({\bf k}'',i\omega_n)
T_{tot}({\bf k}'',{\bf k}',i\omega_n),
\end{equation}
where $u({\bf k},{\bf k}')=\rho_3 v+i\tau_3\Delta g\, v
[{\bf \hat{k}}\times{\bf \hat{k}}']_z$ is the impurity potential
including the spin-orbit contribution. Because of the
angular dependence of the spin-orbit interaction, it can be shown that
the t-matrix (\ref{tmatrix1}) reduces to
$T_{tot}({\bf k},{\bf k}'i\omega_n)=
T(i\omega_n)+T_{so}({\bf k},{\bf k}',i\omega_n)$ where 
\begin{equation}
\label{tn}
T(i\omega_n)=\rho_3 v+\rho_3 v\sum_{{\bf k}}G({\bf k},i\omega_n) T(i\omega_n)
\end{equation} is the usual
momentum-independent contribution from non-magnetic impurities \cite{hirsch1}
and
\begin{equation}
\label{tmatrix2}
T_{so}({\bf k},{\bf k}',i\omega_n)=i\tau_3\Delta g\, v
[{\bf \hat{k}}\times{\bf \hat{k}}']_z+
i\Delta g v\sum_{{\bf k}''}[{\bf \hat{k}}\times{\bf \hat{k}}'']_z
\tau_3 G({\bf k}'',i\omega_n)T_{so}({\bf k}'',{\bf k}',i\omega_n),
\end{equation}
is the t-matrix for the spin-orbit interaction.
Before proceeding with the complete solution of eq.(\ref{tmatrix2}), it is
useful to analyze the lowest order contributions in $\Delta g$.
By replacing $G({\bf k},i\omega_n)$ with $G_0({\bf k},i\omega_n)$, 
the expansion of 
eq.(\ref{tmatrix2}) up to the third order in $\Delta g$ leads to
a spin-orbit part of the self-energy of the form 
$\Sigma_{so}({\bf k},i\omega_n)
=\Sigma_{so}^{(2)}({\bf k},i\omega_n)+
\Sigma_{so}^{(3)}({\bf k},i\omega_n)$,
where the first term is the usual Born contribution
$\Sigma_{so}^{(2)}({\bf k},i\omega_n)=n_i\Delta g^2 v^2
\sum_{{\bf k}'}|{\bf \hat{k}}\times{\bf \hat{k}}'|^2
\tau_3 G_0({\bf k}',i\omega_n)\tau_3$ which renormalizes both the
frequency and, contrary to the normal impurity scattering, the gap
function. The term proportional to $\Delta g^3$ is instead:
\begin{eqnarray}
\label{third}
\Sigma_{so}^{(3)}({\bf k},i\omega_n)&=&-i n_i (\Delta g\,v)^3
\sum_{{\bf k}',{\bf k}''}[{\bf \hat{k}}\times{\bf \hat{k}}']_z
[{\bf \hat{k}}'\times{\bf \hat{k}}'']_z[{\bf \hat{k}}''\times{\bf \hat{k}}]_z
\tau_3G_0({\bf k}',i\omega_n)\tau_3G_0({\bf k}'',i\omega_n)\tau_3 \nonumber \\
&=&-i2\sin(2\phi)\Gamma\frac{\Delta g^3}{c^3}
\left\langle\frac{i\omega_n\sin(\phi')^2}{\sqrt{\Delta(\phi')^2+
\omega_n^2}}\right\rangle
\left\langle\frac{\sin(\phi')^2\Delta(\phi')}{\sqrt{\Delta(\phi')^2+
\omega_n^2}}\right\rangle\rho_2\tau_2\tau_3,
\end{eqnarray}
which contributes to the off-diagonal part of the self-energy.
This term is proportional to $i\sin(2\phi)$ and has therefore a $id_{xy}$
symmetry which is a consequence of the angular dependence of the spin-orbit
potential. Note that eq.(\ref{third}) closely resembles the off-diagonal
contribution found by Balatsky in the context of spin-orbit coupling
of the condensate to magnetic scattering centres \cite{balatsky}.
Although as a function of $\phi$, 
$\Sigma_{so}^{(3)}({\bf k},i\omega_n)$ has maximum contribution where
the gap function $\Delta\cos(2\phi)$ vanishes, it does not open any
gap in the excitation spectrum. In fact, after analytical continuation
$i\omega_n\rightarrow \omega$, it can be shown that at the nodes
of $\Delta\cos(2\phi)$, $\phi=\pm\pi/4$,$\pm 3\pi/4$, the real part 
of the pole of the Green's function for ${\bf k}={\bf k}_F$
satisfies $\omega=\Sigma_{so}^{(3)}({\bf k}_F,\omega)\propto
\Gamma(\Delta g/c)^3(\omega/2\Delta)\log(2\Delta/|\omega|)$, so that
$\omega=0$ is the solution and no additional gap is opened.

Equation (\ref{third}) suggests that the Green's function solution of
the t-matrix problem is of the form:
\begin{equation}
\label{green2}
G({\bf k},i\omega_n)^{-1}=i\tilde{\omega}_n-\rho_3\tilde{\epsilon}_n
-\rho_2\tau_2[\tilde{\Delta}_n(\phi)+i\tau_3\tilde{\Omega}_n(\phi)],
\end{equation} 
where $\tilde{\omega}_n$, $\tilde{\epsilon}_n$, $\tilde{\Delta}_n(\phi)$
and $\tilde{\Omega}_n(\phi)$ are frequency dependent quantities
which must be calculated self-consistently. Note that
in eq.(\ref{green2}), $\tilde{\Omega}_n(\phi)$ is the off-diagonal contribution
which reduces to eq.(\ref{third}) at the lowest order in $\Delta g$.
The t-matrix for the spin-orbit interaction (\ref{tmatrix2}) is solved
by setting $T_{so}({\bf k},{\bf k}',i\omega_n)=i\Delta g\,v[{\bf \hat{k}}
\times{\bf t}({\bf \hat{k}}',i\omega_n)]_z\tau_3$, where
\begin{equation}
\label{t}
{\bf t}({\bf \hat{k}}',i\omega_n)={\bf \hat{ k}}'+
i\Delta g\,v\sum_{{\bf k}''}{\bf \hat{k}}''\tau_3G({\bf k}'',i\omega_n)
[{\bf \hat{k}}''\times{\bf t}({\bf \hat{k}}',i\omega_n)]_z .
\end{equation}
The above equation is actually a system of two coupled equations for the
components $t_x({\bf \hat{k}}',i\omega_n)$ and $t_y({\bf \hat{k}}',i\omega_n)$
which can be explicitly solved and, after some algebra, the spin-orbit 
contribution 
$T_{so}({\bf k},{\bf k},i\omega_n)$ to the total t-matrix
becomes:
\begin{eqnarray}
\label{tmatrix3}
T_{so}({\bf k},{\bf k},i\omega_n)&=&i\Delta g\,v \,
\hat{k}_x\hat{k}_y[A_{yx}^{-1}(i\omega_n)-
A_{xy}^{-1}(i\omega_n)]\tau_3 \nonumber \\
&+&(\Delta g\,v)^2\sum_{{\bf k}'}
\left[A_{yx}^{-1}(i\omega_n)(\hat{k}_x\hat{k}_y')^2 
+A_{xy}^{-1}(i\omega_n)(\hat{k}_y\hat{k}_x')^2\right]
\tau_3G({\bf k}',i\omega_n)\tau_3 ,
\end{eqnarray}
where $A_{xy}^{-1}(i\omega_n)$ and $A_{yx}^{-1}(i\omega_n)$ 
are the inverse of the $4\times 4$
matrices
\begin{eqnarray}
\label{axy}
A_{xy}(i\omega_n)&=&1-(\Delta g\,v)^2\left
[\sum_{{\bf k}}(\hat{k}_x)^2\tau_3G({\bf k},i\omega_n)
\right]\left[\sum_{{\bf k}}(\hat{k}_y)^2\tau_3G({\bf k},i\omega_n)\right] , \\
\label{ayx}
A_{yx}(i\omega_n)&=&1-(\Delta g\,v)^2\left
[\sum_{{\bf k}}(\hat{k}_y)^2\tau_3G({\bf k},i\omega_n)
\right]\left[\sum_{{\bf k}}(\hat{k}_x)^2\tau_3G({\bf k},i\omega_n)\right] .
\end{eqnarray}
Note that the first term in the right hand side of eq.(\ref{tmatrix3})
is proportional to $\hat{k}_x\hat{k}_y=(1/2)\sin(2\phi)$ and, in fact, it
reduces to eq.(\ref{third}) at the lowest order in $\Delta g$.
Moreover, this term would absent be for $s$-wave superconductors since
in this case $A_{xy}(i\omega_n)=A_{yx}(i\omega_n)$.

Plugging equation (\ref{green2}) into  (\ref{axy})
and (\ref{ayx}) permits to invert $A_{xy}(i\omega_n)$ and 
$A_{yx}(i\omega_n)$, and the final 
equations for  $\tilde{\omega}_n$, $\tilde{\epsilon}_n$, 
$\tilde{\Delta}_n(\phi)$
and $\tilde{\Omega}_n(\phi)$ are obtained by demanding self-consistency for
equations (\ref{green1}), (\ref{tn}), (\ref{green2}), 
(\ref{tmatrix3}-\ref{ayx}). A consistent solution requires that
$\tilde{\Delta}_n(\phi)=\tilde{\Delta}_n\cos(2\phi)$ and 
$\tilde{\Omega}_n(\phi)=\tilde{\Omega}_n\sin(2\phi)$, where
\begin{eqnarray}
\label{delta}
\tilde{\Delta}_n&=&\Delta+\Gamma\frac{(c/\Delta g)^2-
(f_n^2-g_n^2)}
{[(c/\Delta g)^2-(g_n^2+f_n^2 )]^2-
4f_n^2g_n^2}f_n ,\\
\label{Omega}
\tilde{\Omega}_n&=&-2\Gamma\frac{f_ng_n(c/\Delta g)}
{[(c/\Delta g)^2-(g_n^2+f_n^2 )]^2-4f_n^2g_n^2},
\end{eqnarray}
where
\begin{equation}
\label{gf}
g_n=\left\langle\frac{i\tilde{\omega}_n\sin(\phi)^2}
{\sqrt{\tilde{\Delta}_n(\phi)^2+\tilde{\Omega}_n(\phi)^2+
\tilde{\omega}_n^2}}\right\rangle 
\,\,,\,\,\,
f_n=\left\langle\frac{\tilde{\Delta}_n(\phi)\sin(\phi)^2}
{\sqrt{\tilde{\Delta}_n(\phi)^2+\tilde{\Omega}_n(\phi)^2+
\tilde{\omega}_n^2}}\right\rangle ,
\end{equation}
while the equation for the renormalized frequency $\tilde{\omega}_n$ is:
\begin{equation}
\label{omegan}
i\tilde{\omega}_n=i\omega_n+\Gamma\frac{g_{0n}}{c^2-g_{0n}^2}+
\Gamma\frac{(c/\Delta g)^2+(f_n^2-g_n^2)}
{[(c/\Delta g)^2-(g_n^2+f_n^2 )]^2-4f_n^2g_n^2}g_n,
\end{equation}
and
\begin{equation}
\label{g0}
g_{0n}=\left\langle\frac{i\tilde{\omega}_n}
{\sqrt{\tilde{\Delta}_n(\phi)^2+\tilde{\Omega}_n(\phi)^2+
\tilde{\omega}_n^2}}\right\rangle .
\end{equation}

The set of equations (\ref{delta})-(\ref{g0}) represents the main result
of this paper and several features can be already outlined. First,
although the equation for the renormalized frequency (\ref{omegan})
is more complex than the simple minded eq.(\ref{renorm2}), 
the conjecture discussed in the introduction is confirmed, i. e.,
as long as $c/\Delta g\ll 1$ the renormalization due to spin-orbit
interaction is of the same order of magnitude of the one induced by
the impurity scattering. In addition, the spin-orbit scattering
renormalizes also the gap function and induces the additional off-diagonal 
self-energy $\tilde{\Omega}_n$. From eq.(\ref{Omega}) it can be seen that
$\tilde{\Omega}_n\propto (\Gamma/\Delta) (\Delta g/c)^3$ for $\Delta g/c\ll 1$
and $\tilde{\Omega}_n\propto (\Gamma/\Delta) c/\Delta g$ for $c/\Delta g\ll 1$
and is therefore much smaller than the spin-orbit parts of $\tilde{\Delta}_n$
and $\tilde{\omega}_n$ which are of order $(\Gamma/\Delta) (\Delta g/c)^2$
for $\Delta g/c\ll 1$ and $\Gamma/\Delta$ for $c/\Delta g\ll 1$, respectively.

Equations (\ref{delta})-(\ref{g0}) must be completed with the
equation for the order parameter $\Delta$ which, if the pairing interaction is
$V_{pair}({\bf k},{\bf k}')=-V_{pair}\cos(2\phi)\cos(2\phi')$, reduces to:
\begin{equation}
\label{orderpar}
\Delta=\frac{V_{pair}}{4} T\sum_n \sum_{{\bf k}}\cos(2\phi)
\mbox{Tr}\left[\rho_2\tau_2 G({\bf k},i\omega_n)\right]=
\lambda \pi T\sum_n \left\langle\frac{\tilde{\Delta}_n(\phi)\cos(2\phi)}
{\sqrt{\tilde{\Delta}_n(\phi)^2+\tilde{\Omega}_n(\phi)^2+
\tilde{\omega}_n^2}}\right\rangle,
\end{equation}
where $\lambda=V_{pair}N_0$ is the coupling constant.
The critical temperature $T_c$ is obtained from eq.(\ref{orderpar})
by setting $\tilde{\Delta}_n\rightarrow 0$ and $\tilde{\Omega}_n\rightarrow 0$
in equations (\ref{delta})-(\ref{orderpar}). The resulting critical
temperature satisfies a Abrikosov-Gorkov type of relation \cite{AG}:
\begin{equation}
\label{ag1}
\log\left(\frac{T_c}{T_{c0}}\right)=
\psi\!\left(\frac{1}{2}\right)-\psi\!\left(\frac{1}{2}+
\frac{\Gamma_n+\Gamma_{so}}{2\pi T_c}\right),
\end{equation}
where $\psi$ is the di-gamma function and $T_{c0}$ is the critical 
temperature for the pure system ($\Gamma=0$). The impurity and
spin-orbit scattering parameters are
\begin{equation}
\label{ag2}
\Gamma_n=\frac{\Gamma}{c^2+1}\,, \,\,\,\,\,\,\,\,
\Gamma_{so}=\Gamma\frac{3(2c/\Delta g)^2+1}{[(2c/\Delta g)^2+1]^2} .
\end{equation}
Superconductivity is destroyed for the critical value
$\Gamma_n^{\star}+\Gamma_{so}^{\star}=\pi T_{c0}/2\gamma$ 
($\gamma\simeq 1.781$). For $\Delta g=0$, the $c\rightarrow 0$ limit
leads to the usual result for scalar impurities 
$\Gamma^{\star}=\pi T_{c0}/2\gamma$ \cite{maki1} while, for $\Delta g\neq 0$,
$\lim_{c\rightarrow 0}\Gamma^{\star}=\pi T_{c0}/4\gamma$,
i. e., half the value of the impurity concentration in the unitarity limit
without spin-orbit scattering. This result clearly
shows that, in the unitarity limit,  spin-orbit scattering has a strong 
pair-breaking effect of the same size of scalar impurity scattering.
This feature is a consequence of the infinite resummation of the t-matrix
which gives rise to a non-analytic point at $c=0$, $\Delta g=0$, and the
limit $\Delta g\rightarrow 0$,$c=0$ differs from
$c\rightarrow 0$,$\Delta g = 0$.
The reduced critical temperature $T_c/T_{c0}$ (solid lines) and the 
reduced zero temperature 
order parameter $\Delta/\Delta_0$ (dashed lines)
are shown in fig. 1 as a function of $\Gamma/\Delta_0$ for $c=0.1$
and $\Delta g =0.0$, $0.05$, $0.1$, $0.15$ and $0.2$ (from right to left). 
$\Delta$ is
calculated numerically from the $T\rightarrow 0$ limit
of eq. (\ref{orderpar}) and $\Delta_0$ is the gap for the pure case (in
the weak coupling limit). As a function of impurity concentration, the 
suppression of both $T_c/T_{c0}$ and
$\Delta/\Delta_0$ is stronger for smaller values of $c/\Delta g$.
In the inset of fig. 1, $T_c/T_{c0}$ and $\Delta/\Delta_0$ are shown for
the limiting cases $c\rightarrow 0$,$\Delta g=0$ (a) and
$\Delta g\rightarrow 0$,$c=0$ (b).

Another thermodynamic quantity of interest is the
superfluid density $\rho_s$ which is calculated from 
\begin{equation}
\rho_s=2\pi m N_0v_F^2 T\sum_m\left\langle
\frac{\tilde{\Delta}_n(\phi)\cos(\phi)^2}
{[\tilde{\Delta}_n(\phi)^2+\tilde{\Omega}_n(\phi)^2+
\tilde{\omega}_n^2]^{3/2}}\right\rangle,
\end{equation}
where $v_F$ and $m$ are the electron Fermi velocity and mass, 
respectively \cite{choi}. In fig. 2 it is shown $T_c/T_{c0}$
as a function of the zero temperature limit of $\rho_s/\rho_{s0}$, 
where $\rho_{s0}$ is the superfluid density for the
pure system, for $\Gamma/\Delta_0=0.1$, $\Delta g=0.1$ and
different values of $c$. Note that, by lowering $c$, the curves move
upwards and that this effect is visible also for the limiting cases
$c\rightarrow 0$,$\Delta g=0$ (a) and
$\Delta g\rightarrow 0$,$c=0$ (b) reported in the inset. This feature
is given by the spin-orbit renormalization of $\tilde{\Delta}_n$, otherwise
absent when $\Delta g=0$.
In Ref. \cite{franz}, it has been stressed that
when $T_c/T_{c0}$ is plotted against the zero temperature limit of
$\rho_s/\rho_{s0}$ the experimental data lay above the theoretical curve 
corresponding to the unitarity limit. 
As inferred from both fig. 2 and the inset, the 
inclusion of spin-orbit scattering tends to cure this discrepancy.

In addition to thermodynamic quantities, the spin-orbit interaction affects
also the spectral properties of $d$-wave superconductors. In fact,
nonzero values of $\Delta g$ contribute to the amount of gapless
excitations already provided by the resonant scattering with impurities. 
This feature can be investigated by performing the analytical continuation
of eqs. (\ref{delta})-(\ref{g0}) by setting $i\omega_n\rightarrow \omega$ and 
$i\tilde{\omega}_n\rightarrow \tilde{\omega}$ and then plugging the
results into the quasiparticle density of states $N(\omega)$ given below:
\begin{equation}
\frac{N(\omega)}{N_0}=\mbox{sgn}(\tilde{\omega})\mbox{Re}
\left\langle\frac{\tilde{\omega}}{\sqrt{\tilde{\omega}^2-
\tilde{\Delta}(\phi)^2-\tilde{\Omega}(\phi)^2}}\right\rangle .
\end{equation}
The resulting density of states is plotted in fig. 3 for $\Gamma/\Delta_0=
0.05$, $c=0.05$ (a) and for $\Gamma/\Delta_0=0.1$, $c=0.1$ (b). 
In both cases, nonzero values of $\Delta g$ provide an additional
contribution to the gapless states lowering at the same time
the intensity of the coherence peaks at $\omega\simeq \Delta$. In
the insets of fig. 3 it is also shown how the zero energy density of states
$N(0)/N_0$ is affected by the spin-orbit coupling. 

In summary, it has been shown that, if the impurity potential is
close to the unitarity limit, the spin-orbit coupling to the impurities
is as important as the scalar impurity potential and its effects 
cannot be neglected. This result points toward a critical re-examination
of the existing experimental data in addition to the recent theoretical
developments based on the spatial variation of the order parameter.
As a final remark, it should be noted that a promising experimental
route for the estimation of the spin-orbit interaction
in $d$-wave superconductors is provided by the Zeeman response under
an external magnetic field \cite{yang,maki2}. Preliminary
results suggest that nonzero values of $\Delta g$
drastically affect the Zeeman response of $d$-wave superconductors
\cite{grima}. 

\stars{The author would like to thank P. Fulde and I. Vobornik
for fruitful discussions.}

\newpage

\begin{figure}
\protect
\centerline{\psfig{figure=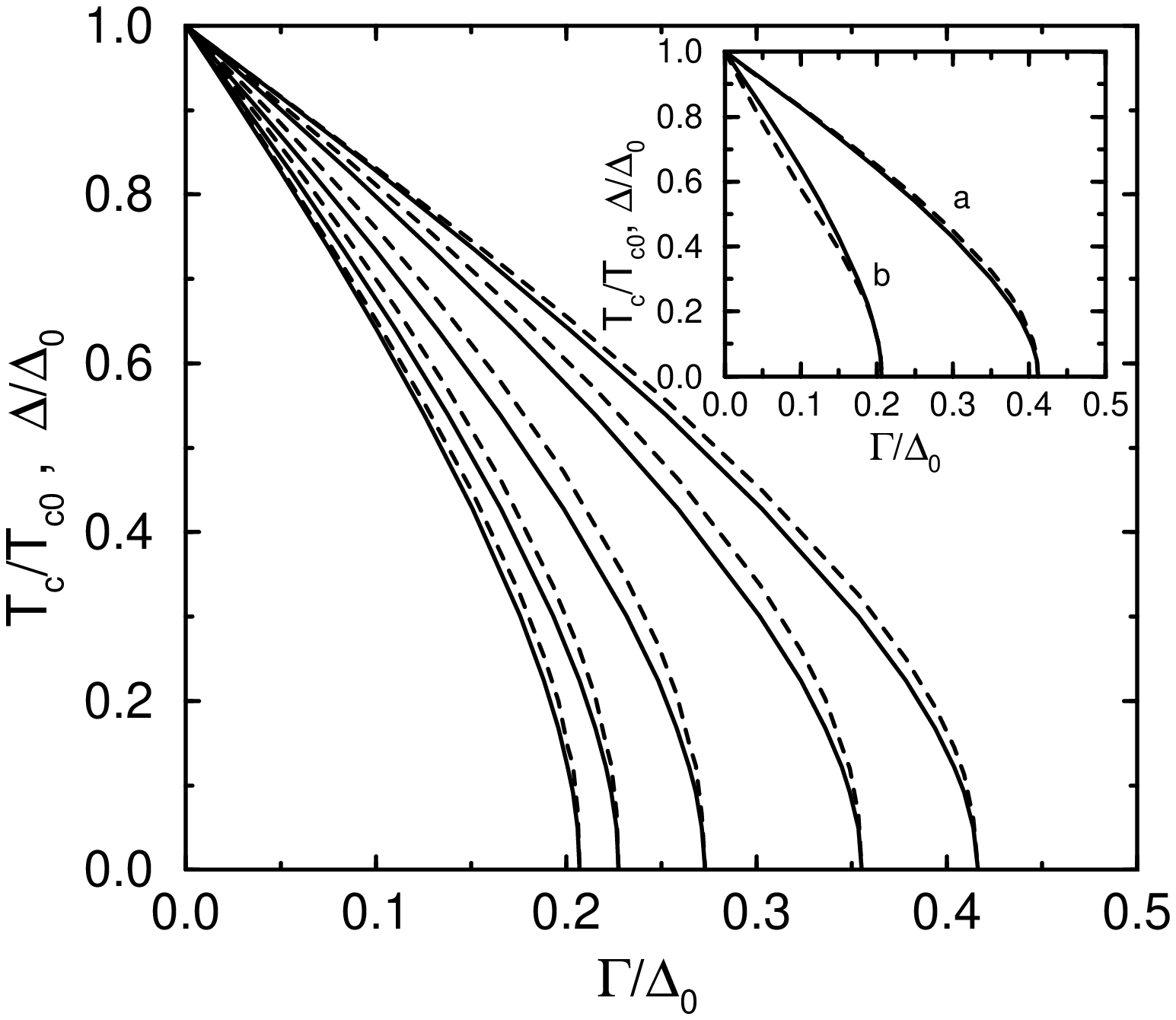,width=10cm}}
\caption{ Reduced critical temperature (solid lines) and order parameter
(dashed lines) as function of  $\Gamma/\Delta_0$ for $c=0.1$ and
$\Delta g=0.0$, $0.05$, $0.1$, $0.15$, and $0.2$ (from right to left).
Inset:  the same quantities for the $c\rightarrow 0$,$\Delta g=0$ limit
(a) and for the $\Delta g\rightarrow 0$,$c=0$ limit (b)}
 \label{fig1}
\end{figure}

\begin{figure}
\protect
\centerline{\psfig{figure=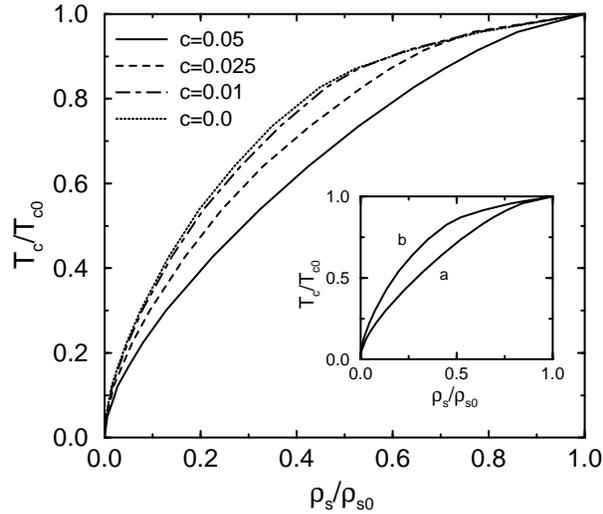,width=10cm}}
\caption{Reduced critical temperature as a function of the
reduced superfluid density for $\Gamma=0.1$, $\Delta g=0.1$ and 
different values of $c$. 
Inset: superfluid density in the $c\rightarrow 0$,$\Delta g=0$ limit
(a) and in the $\Delta g\rightarrow 0$,$c=0$ limit (b)}
\label{fiag2}
\end{figure}

\begin{figure}
\protect
\centerline{\psfig{figure=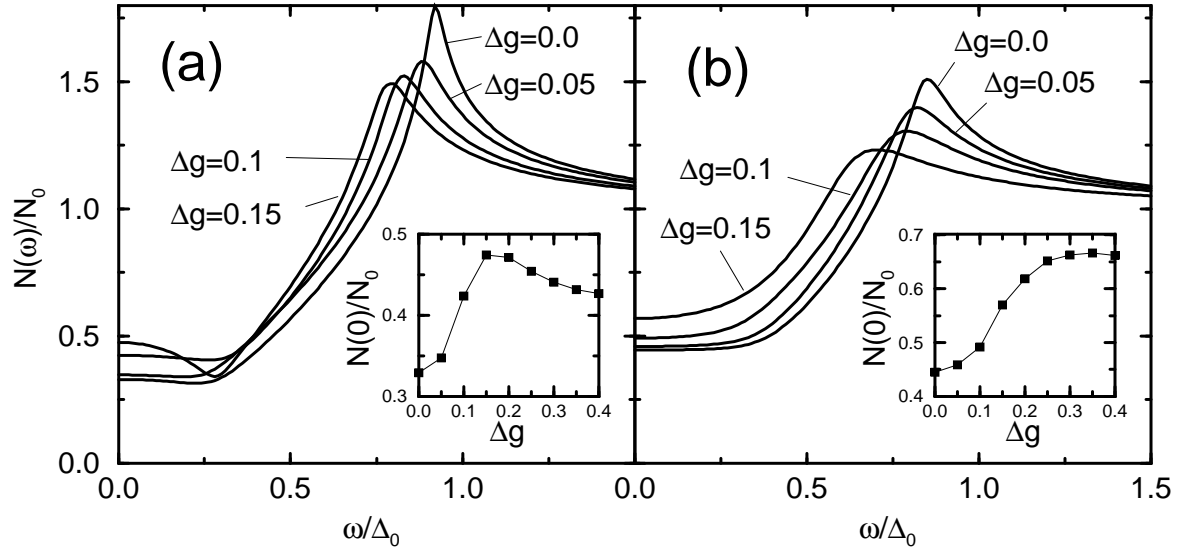,width=15cm}}
\caption{Quasiparticle density of states for $\Gamma=0.05$, $c=0.05$ (a)
and $\Gamma=0.1$, $c=0.1$ (b) and different values of $\Delta g$.
Insets: corresponding zero energy density of states as a function of $\Delta g$. }
\label{fig3}
\end{figure}

\end{document}